\documentclass[preprint]{vgtc}               

\graphicspath{{figures/}{pictures/}{images/}{./}} 

\usepackage{times}                     

\usepackage{tabu}                      
\usepackage{booktabs}                  
\usepackage{lipsum}                    
\usepackage{mwe}                       

\usepackage{mathptmx}                  

\usepackage{amsmath}
\usepackage{amsfonts}
\usepackage{enumitem}
\usepackage[svgnames]{xcolor}

\newcommand{\metricitem}[1]{\vspace{0.5em}\noindent\textbf{#1}~}

\onlineid{0}

\vgtccategory{Research}

\vgtcinsertpkg

\preprinttext{To appear in IEEE International Symposium on Mixed and Augmented Reality.}

\title{An Embodied AR Navigation Agent: Integrating BIM with Retrieval-Augmented Generation for Language Guidance}

\author{
  Hsuan-Kung Yang\thanks{Author e-mails: \{hsuan-kung.yang, tsuching.hsiao, ryoichiro.oka, ryuya.nishino, satoko.tofukuji, norimasa.kobori\}@woven.toyota} \and 
  Tsu-Ching Hsiao\footnotemark[1] \and 
  Ryoichiro Oka\footnotemark[1] \and 
  Ryuya Nishino\footnotemark[1] \and 
  Satoko Tofukuji\footnotemark[1] \and 
  Norimasa Kobori\footnotemark[1] \and
}

\affiliation{
  Woven by Toyota, Inc., Japan
}

\teaser{
  \centering
  \includegraphics[width=\linewidth]{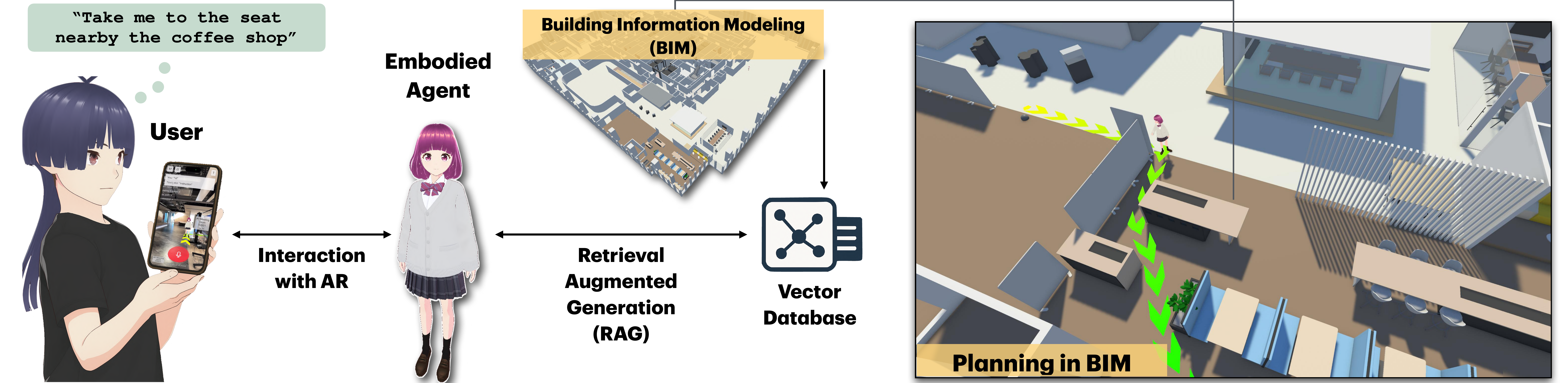}
  \caption{The proposed AR navigation system integrates a multi-agent RAG framework with BIM data to support flexible natural language queries for goal retrieval, interaction, and navigation. Communication is enabled through an AR embodied agent interface.}
  \label{fig:teaser}
}

\abstract{
Delivering intelligent and adaptive navigation assistance in augmented reality (AR) requires more than visual cues, as it demands systems capable of interpreting flexible user intent and reasoning over both spatial and semantic context. Prior AR navigation systems often rely on rigid input schemes or predefined commands, which limit the utility of rich building data and hinder natural interaction. In this work, we propose an embodied AR navigation system that integrates Building Information Modeling (BIM) with a multi-agent retrieval-augmented generation (RAG) framework to support flexible, language-driven goal retrieval and route planning. The system orchestrates three language agents, Triage, Search, and Response, built on large language models (LLMs), which enables robust interpretation of open-ended queries and spatial reasoning using BIM data. Navigation guidance is delivered through an embodied AR agent, equipped with voice interaction and locomotion, to enhance user experience. A real-world user study yields a System Usability Scale (SUS) score of 80.5, indicating excellent usability, and comparative evaluations show that the embodied interface can significantly improves users’ perception of system intelligence. These results underscore the importance and potential of language-grounded reasoning and embodiment in the design of user-centered AR navigation systems. Video demonstrations are available at \href{https://woven-visionai.github.io/ar-navigation-agent-project}{https://woven-visionai.github.io/ar-navigation-agent-project}.
} 


\begin{document}


\firstsection{Introduction}

\maketitle

Augmented Reality (AR) has emerged as a compelling interface for navigation, which offers intuitive and real-time visual guidance by overlaying virtual elements onto real-world environments~\cite{narzt2006augmented,bhorkar2017survey}. Early AR navigation systems primarily relied on directional arrows or path lines to indicate routes~\cite{bhorkar2017survey}. While such methods are simple and straightforward to implement, they often introduce perceptual ambiguities, particularly in spatially complex environments where users may misjudge direction or depth when following floating indicators. To address these limitations, more expressive forms of guidance have been explored, such as the use of embodied agents or avatars that mimic human behavior to guide users along a planned path~\cite{murata2014ar}. These human-like guides not only improve clarity but also offer a sense of presence and reassurance during navigation, thereby extending the potential for engaging interaction.
 
While AR interfaces have become increasingly expressive, the integration of spatial and semantic understanding remains a core challenge for delivering context-aware guidance in real-world environments. Effective navigation requires not only accurate localization and path planning, but also the ability to reason about the semantic properties of surrounding spaces, such as room functions, accessibility, or proximity to relevant landmarks. However, most AR systems struggle to unify these two dimensions. Building Information Modeling (BIM) has emerged as a promising solution to this problem which offers structured representations of both geometry and semantic metadata within environments~\cite{kensek2014building}. Specifically, BIM data includes detailed 3D layouts at accurate geometric scales, annotated room identities, and infrastructure attributes, all of which can be leveraged for both route planning and augmented visualization. When combined with accurate localization, this enables AR systems to render AR guidance with precise alignment to the real-world environment, while also incorporating contextual information to support more context-oriented and adaptive navigation.

Despite advancements in integrating spatial grounding with semantic context, as exemplified by systems that leverage BIM, many AR navigation systems still fall short in how users express their intended destinations and rely on limited input mechanisms for goal specification. Most existing systems require users to select from predefined destination lists, use fixed command phrases~\cite{hou2021ar,hertel2023welcome}, or rely on simple keyword matching~\cite{zhao2020voice}, which restricts the flexibility and expressiveness of user interaction. These constraints prevent the system from fully utilizing the rich spatial-semantic information available in structured models such as BIM. For example, users are often unable to describe goals based on contextual attributes (e.g., “the nearest meeting room with over 8 seats” or “the nearest accessible restroom”), which hinders the system’s ability to reason over multiple candidate locations and select the most appropriate one. As a result, the integration of rich spatial-semantic representations into AR navigation systems remains largely underexploited, limiting the ability of these systems to intelligently interpret user intent and deliver context-aware guidance.

To address this gap, this work proposes a multi-agent retrieval-augmented generation (RAG) framework for goal retrieval, which enables users to interact with an AR navigation system through flexible and natural language input. Rather than relying on fixed commands or predefined  categories or keywords, our approach interprets open-ended user queries and retrieves relevant navigation targets by reasoning over both spatial and semantic information encoded in BIM data. The system is orchestrated by three specialized agents: a \textit{Triage Agent} that classifies user intent and delegates tasks, a \textit{Search Agent} that performs semantic retrieval from a vector database of annotated locations, and a \textit{Response Agent} that synthesizes contextually appropriate responses. These agents are implemented using large language models (LLMs) with RAG capabilities. This multi-agent architecture supports robust and interpretable interactions, allowing the AR system to respond intelligently to a wide variety of user inputs, including vague or descriptive language. By integrating this language-guided reasoning pipeline with BIM-based spatial awareness, the proposed system bridges the gap between semantic query understanding and grounded navigation in real-world environments. For example, if a user says, “I’m hungry, could you suggest somewhere to go?”, the system is able to infer the user’s intent, identify a relevant nearby location (e.g., a coffee shop), plan a route using BIM data, and provide intuitive guidance. The response and navigation cues are presented through an embodied AR agent, offering both verbal feedback and visually grounded assistance for a more natural and engaging user experience. 

To evaluate the effectiveness and usability of the proposed system, we design a user study to investigate two core research questions. First, which interface modality best complements the intelligence and flexibility of the navigation system, as reflected in user perceptions of system intelligence. Second, how usable the proposed multi-agent and RAG-based AR navigation system is when experiences through the embodied agent interface. The study is conducted in a real-world indoor environment and compares two AR guidance modalities: a baseline arrow-only scheme and the full system with an embodied agent complemented by directional arrows. Both interfaces are equipped with the proposed RAG framework for goal retrieval. To address these questions, we measure a range of user experience metrics and incorporated the System Usability Scale (SUS) to assess overall system usability~\cite{bangor2009determining, lewis2018item}. In addition, we present a comprehensive set of qualitative case studies to  examine how the system handles a variety of user queries, ranging from direct destination requests to ambiguous or contextual prompts. These case studies highlight both the advantages of flexible, language-based interaction and the limitations of the  implementation. Collectively, these evaluations provide a holistic understanding of the system’s capabilities and help inform future directions for the development of intelligent AR navigation systems.

The primary contributions of this work are as follows:
\begin{itemize}
  \item To our knowledge, this is the first AR navigation system that integrates a multi-agent RAG framework with BIM data to support language-driven goal retrieval, which enables flexible language interaction grounded in spatial-semantic context.
  \item We demonstrated that an embodied agent equipped with voice interaction and locomotion can significantly enhance users’ perception of system intelligence when interacting with a RAG-based system, compared to delivering the same reasoning capabilities through a non-embodied interface. This finding underscores the critical role of embodiment in shaping user experience when integrating AR navigation with RAG-based reasoning systems.
\end{itemize}

\section{Related Work}
\subsection{AR Navigation System and Interface Design}
Augmented Reality (AR) has long been studied as a medium for spatial navigation~\cite{bhorkar2017survey,Lynam28112024,Cheliotis04052023}, gradually evolving from systems with rigid goal specification to more natural and flexible language interfaces. Early AR navigation systems typically constrained users to select destinations from predefined lists or required exact keyword inputs. For instance, some works~\cite{qin2013campus,wu2018path} developed systems where users selected their destinations through dropdown menus or fixed options. These rigid input schemes limited the expressiveness of user queries and the adaptability of the system. As natural language understanding  progressed, systems such as that of Zhao et al.~\cite{zhao2020voice} introduced voice interaction, allowing users to specify destinations through spoken language. While these systems still required structured phrasing, they marked a significant step toward more intuitive interaction. More recently, the emergence of LLMs~\cite{devlin2018bert, openai2023gpt4, deepseekai2024deepseekv3technicalreport} has enabled systems to interpret open-ended user queries and infer user intent from vague or context-rich input (e.g., “I need a place to relax before my meeting”), expanding the range of possible interactions toward a more human-centric navigation paradigm.

The way AR systems present guidance has also undergone substantial evolution. Traditional systems generally overlay navigation cues like arrows, lines, or path indicators directly onto the live camera view~\cite{bhorkar2017survey,qin2013campus,wu2018path,petr2024reliability,juile2024arguide}. Although such visual cues are simple and effective in controlled or open spaces, they can become ambiguous in visually complex or cluttered environments, where depth perception and occlusion make interpretation challenging~\cite{murata2014ar}. To address these limitations, researchers have explored embodied guidance using virtual agents. Murata et al.~\cite{murata2014ar} introduced a CG avatar that leads users by walking and gesturing along the navigation path, while Kuwahara et al.~\cite{kuwahara2019campus} demonstrated that avatars designed as familiar mascots can improve users’ situational awareness. Virtual agents provide richer, human-like cues, such as pointing, gazing, or waiting, that arrows cannot convey. However, their effectiveness depends heavily on precise tracking and spatial consistency; as Lee et al.~\cite{lee2022mixed} found, users may prefer arrows for their simplicity and clarity in task-driven contexts, despite agents being more engaging. Together, these findings highlight a trade-off between expressiveness and reliability in AR guidance design.

\begin{figure*}[t]
  \centering 
  \includegraphics[width=\textwidth]{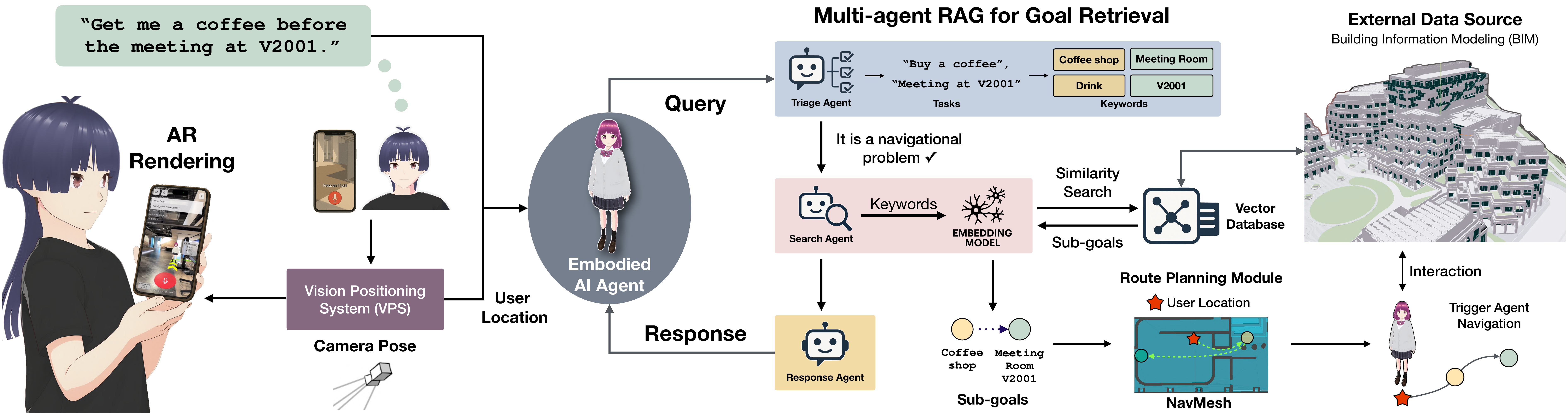}
  \caption{An overview of the proposed AR navigation system which integrates a multi-agent retrieval-augmented generation (RAG) framework with BIM data to support flexible natural language queries for navigation. The proposed system features a multi-agent orchestration architecture composed of three agents: (a) the \textbf{\textit{Triage Agent}} for classifying user queries and delegating tasks accordingly; (b) the \textbf{\textit{Search Agent}} for retrieving relevant destinations and contextual information from a vector database with user queries; and (c) the \textbf{\textit{Response Agent}} for generating appropriate and user-friendly responses based on the aggregated context. Communication between the user and the system is facilitated through an AR embodied agent interface, equipped with text-to-speech (TTS), speech-to-text (STT), and locomotion capabilities.} 
  \label{fig:overview}
\end{figure*}

\subsection{Building Information Modeling for Navigation}
Recent advancements in indoor navigation increasingly rely on semantic understanding of 3D environments, where spatial geometry is tightly coupled with meaningful semantic labels. Scene reconstruction and segmentation with open-vocabulary models, such as CLIP~\cite{radford2021learning}, has emerged as a promising direction, enabling the generation of spatially and semantically aligned maps without the need for manual annotation~\cite{takmaz2023openmask3d, boudjoghra2025openyolo}. These models can flexibly label objects such as “chair,” “table,” or “plant,” making them suitable for general-purpose scene parsing and allowing users to search for objects within a 3D scene using natural language queries~\cite{xu2024multimodal}.
However, they often lack higher-level functional semantics, such as “exhibition area,” “toilet,” or “coffee shop,” as well as fine-grained attributes such, as “room size” and “seating capacity,” which are essential for natural and goal-directed navigation in real-world environments. In contrast, BIM and Industry Foundation Classes (IFC) provide structured representations that incorporate both geometric layouts and human-curated semantic information about the function and usage of indoor spaces. This rich contextual grounding makes BIM and IFC a more suitable foundation for navigation systems that require accurate interpretation of user intent~\cite{liu2021indoor}. Previous work has demonstrated how BIM and IFC can be used to support spatial route planning for robots~\cite{gopee2023improving, torres2023ogm2pgbm,karimi2021semantic}, deliver AR-based guidance for human users~\cite{zhang2025bim,tang2025bim}, and facilitate emergency navigation~\cite{valizadeh2024indoor,ahn2024bim}. 

Recent research in BIM information retrieval has explored the integration of LLMs with spatial and semantic indexing techniques. Prior work~\cite{bim-gpt} proposes a system capable of answering natural language queries over BIM data, while another study~\cite{graph-rag-ifc} incorporates Graph-RAG to enable relation-aware and spatially enriched retrieval. Contemporary studies~\cite{multi-agent-bim, spatial-tree} further employ multi-agent frameworks to resolve complex spatial relationships within BIM environments. Nevertheless, these methods remain limited in their ability to incorporate user information in real-time. In contrast, the proposed system introduces a multi-agent RAG architecture with a two-step retrieval mechanism that considers user location and spatial information to enhance the contextual retrieval. This mechanism enables seamless integration with an AR-based navigation system and supports real-time spatial guidance, thereby bridging the gap between information retrieval and in-situ task-oriented navigation, an aspect not addressed in prior approaches.

\subsection{Large Language Model and Embodied AI Agent}
Large Language Models (LLMs)~\cite{devlin2018bert,openai2023gpt4}, such as OpenAI’s GPT-4~\cite{openai2023gpt4} and DeepSeek~\cite{deepseekai2024deepseekv3technicalreport}, have shown impressive general-purpose capabilities, including natural dialogue, reasoning, and context understanding. Their versatility has driven the development of applications that assist users in solving everyday problems through conversational interfaces, ranging from productivity tools~\cite{peng2023impact} to educational assistants~\cite{wang2024large} and healthcare applications~\cite{he2025survey}. These models are increasingly being integrated with extended reality (XR) and robotics systems to enable more human-like, intelligent interactions with the physical world, such as controlling robots~\cite{cheng2024navila} and smart home devices~\cite{king2024sasha} using flexible natural language queries, or providing intuitive spatial instructions via XR to assist with real-world problem solving~\cite{srinidhi2024xair}. 

As LLMs became more advanced, researchers began exploring ways to bring them into physical and virtual environments through embodied agents~\cite{yang2025embodied}, aiming to create more immersive and engaging user experiences. Recent studies have introduced anthropomorphized objects~\cite{nagano2023talk, iwai2025bringing}, where items are enhanced with voices and personalities to support conversational interaction, as well as human-shaped avatars~\cite{reinhardt2020embedding, yang2024effects} that communicate through natural language and expressive behavior. Some works also integrated AI-generated animation to make agent behavior more natural~\cite{gunawardhana2024toward}. These studies often focus on user preferences for different embodiments and investigate how the visual form of an agent affects perceived intelligence, trustworthiness, and social presence~\cite{skarbez2017survey}. Such embodied agents have been widely applied in cultural and educational contexts, particularly as conversational guides in virtual museums~\cite{garcia2024speaking, chen2025exploring, rzayev2019effect}, where they can answer visitor questions and provide navigational support within the virtual world~\cite{cao2021interactive}, enhancing both engagement and spatial understanding. This combination of LLMs and embodied agents represents a broader shift toward more intuitive, natural human-agent interaction. However, the systematic integration of embodied agents with AR-based navigation applications, which are capable of delivering real-world guidance through flexible language instructions, remains largely unexplored, leaving a gap in bridging embodied AI with physical space interaction.

\section{Multi-Agent RAG System for Navigation}
\label{sec:multi-agent-rag}
\subsection{Overview and Problem Formulation}
\begin{figure}[t]
  \centering 
  \includegraphics[width=.99\linewidth]{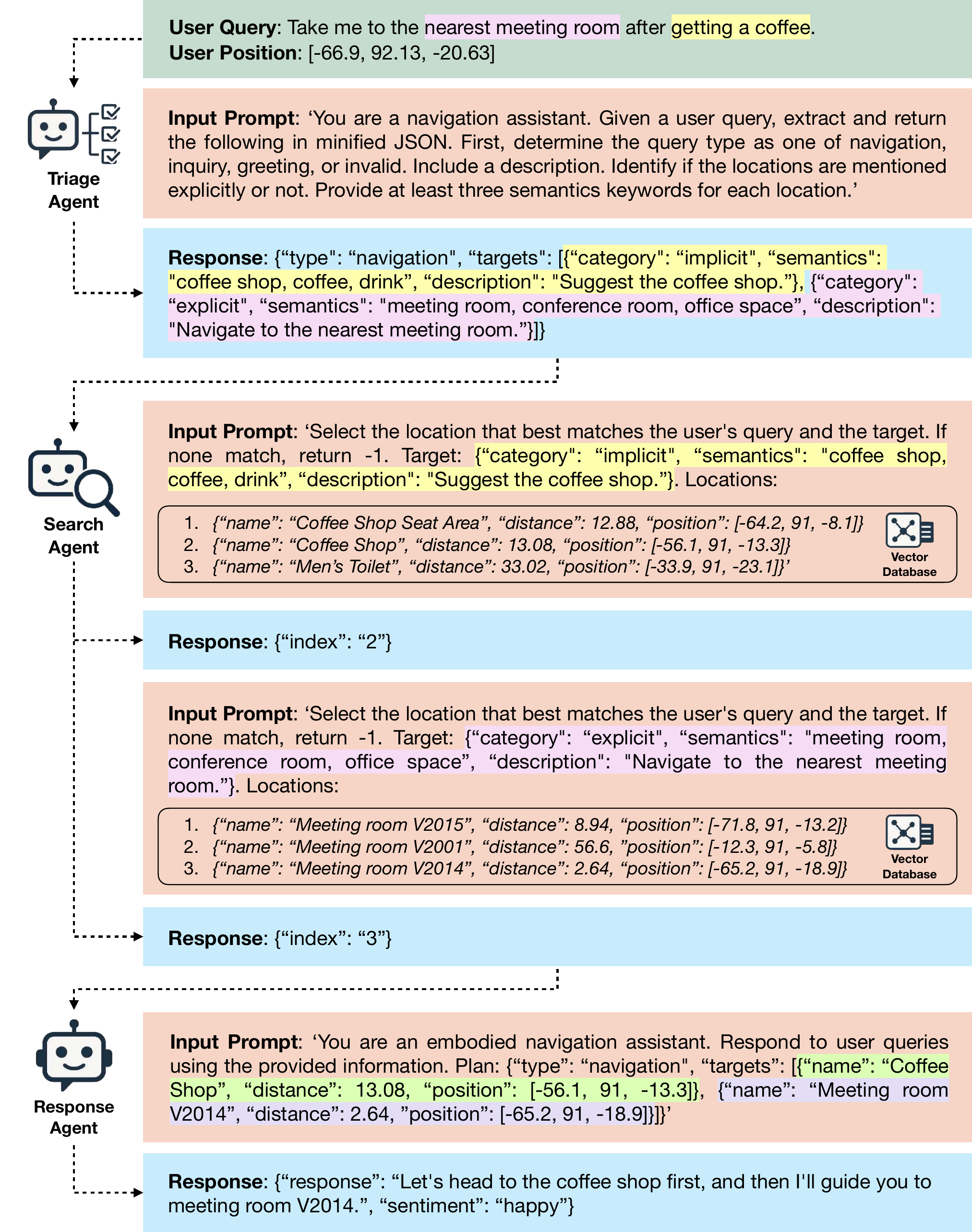}
  \caption{
    Example workflow of the multi-agent RAG system for navigation. Given a user query and the user’s current position, the \textit{Triage Agent} classifies the query as a navigation task, extracts semantic targets (e.g., “getting a coffee”), and generates corresponding keyword sets (e.g., “coffee shop, coffee, drink”) in the user-specified navigation order. These keyword sets are passed to the \textit{Search Agent}, which encodes them using a sentence encoder and retrieves the top N candidates from the vector database based on cosine similarity. The retrieved candidates, along with the semantic target, are appended to the prompt, and the agent selects the most relevant location that best matches the user query and intent. If the query includes spatial qualifiers such as “nearest”, the \textit{Search Agent} also considers distance, computed from the user’s current position, during selection. The chosen location is then passed to the \textit{Response Agent}, which generates natural language navigation instructions grounded in both semantic and spatial context.
  }
  \label{fig:rag-sample}
\end{figure}

Given the user's navigation query \(Q_t\) at time step \(t\), such as requesting guidance to a specific location (e.g., ``Get me a coffee before the meeting at V2001''), our objective is to accurately identify and retrieve goal-related spatial information and subsequently deliver intuitive navigational guidance. The proposed framework employs a multi-agent retrieval-augmented generation (RAG) architecture to process these queries. Initially, the \textit{Triage Agent} classifies the user input, extracts semantic keywords, and formulates structured target necessary for subsequent retrieval processes. The obtained metadata is then utilized by the \textit{Search Agent} to perform semantic similarity searches within a high-dimensional vector database, retrieve relevant spatial data from an external BIM source, and decide the goal. Concurrently, the input frame $I_t$ from the user's device is processed by a Vision Positioning System (VPS) to estimate camera orientation and position. This enables accurate AR rendering, which is essential for integrating virtual navigational cues into the physical environment.
The results generated by the RAG system, including both the goal information and the response, are subsequently conveyed through an embodied AI agent equipped with text-to-speech functionality. The goal and the user’s current position are processed by the route planning module to compute the optimal path, which is then sent to the agent for guiding the user. Building on this information, the agent serves as an interactive navigator, directing the user toward specific spatial targets using BIM-derived spatial information and VPS-driven localization. An overview of the proposed system architecture is shown in \cref{fig:overview}, and a representative example of the multi-agent RAG query processing workflow is provided in \cref{fig:rag-sample}.

\subsection{BIM and IFC Data Stream to Vector Database}
\label{subsec:bim-ifc}
To enable efficient semantic retrieval for navigation tasks, we preprocess BIM data by transforming raw IFC data~\cite{buildingSMART2024} into a structured vector database~\cite{2021milvus}. The workflow begins by parsing the IFC data to extract key architectural and spatial entities, including floors, areas, doors, and furniture, such as \texttt{IfcSpace}, \texttt{IfcDoor}, and \texttt{IfcFurnishingElement}. Specifically, each entity is associated with structured metadata $\mathbf{m}_i$ and is formulated as below,
\begin{equation}
    \mathbf{m}_i = \{ \texttt{name}_i,\ \texttt{description}_i,\ \mathbf{p}_i\},
\end{equation}
where $i$ is the index and $\mathbf{p}_i \in \mathbb{R}^3 $ denotes the 3D position of the element in the BIM coordinate system. For elements lacking an explicitly defined pivot point in the IFC model, which is commonly the case for aggregated structures such as floors or open areas, we compute the centroid of their axis-aligned bounding box derived from the 3D geometry and adopt this center point as the representative spatial location. We then generate a descriptive sentence by concatenating the element’s name and description, which includes semantic information (e.g., function, seating capacity, and amenities). These sentences are encoded using a sentence-level transformer model~\cite{reimers-2019-sentence-bert} into dense embeddings,
\begin{equation}
    \mathbf{v}_i = \texttt{Encoder}(\texttt{name}_i  \Vert \texttt{description}_i),
\end{equation}
where $\mathbf{v}_i \in \mathbb{R}^d$ is the resulting embedding vector, $d = 768$ denotes the dimensionality of the embedding, and $\Vert$ denotes the string concatenation. Each embedding $\mathbf{v}_i$ is stored with its associated metadata in a vector database as an entry $\mathcal{E}_i = (\mathbf{v}_i, \mathbf{m}_i)$. Finally, we insert these tuples into a high-dimensional vector database, which supports efficient semantic similarity search. This preprocessing stage lays the foundation for accurate and context-aware goal retrieval in our navigation system. 

\subsection{Multi-Agent Orchestration for Goal Retrieval}
\label{subsec:multi-agent-rag}
To leverage the constructed database, which stores goals along with their semantic attributes and descriptions, we introduce a multi-agent RAG framework designed for goal retrieval. This framework employs an orchestration scheme to delegate tasks among specialized agents. The agents are designed to collaborate with each other and utilize the external knowledge retrieved from the structured vector database, which contains the spatial and semantic data from BIM sources, to perform the navigation and conversation tasks. Specifically, the framework employs three specialized agents: \textit{Triage Agent}, \textit{Search Agent}, and \textit{Response Agent}, that coordinate tasks to efficiently handle user queries involving navigation and facility information. \Cref{fig:rag-sample} illustrates an example of the multi-agent orchestration process, and the roles and responsibilities of each agent are described in the following paragraphs, 

\paragraph{\textbf{Triage Agent.}} 
The \textit{Triage Agent} is designed to classify user queries and initiates multi-agent orchestration by generating structured representations that drive downstream execution. The agent first classifies each query into one of four categories: navigation, inquiry, greeting, or invalid. For navigation-related queries, the agent extracts a structured target $\tau$, which is defined as,
\begin{equation}
    \tau = \{\texttt{category},\; \texttt{semantics},\; \texttt{description}\},
\end{equation}
where \texttt{category} specifies whether the target is \texttt{explicit} (e.g., a named location like ''Meeting Room V2014'') or \texttt{implicit} (e.g., ''a place to get coffee''), \texttt{semantics} is a set of keywords (e.g., ``coffee shop, drink''), and \texttt{description} is a natural language summary of the goal. These components form the basis for semantic retrieval and are passed along with the user’s location to the \textit{Search Agent} for goal retrieval. If the query involves multiple sub-goals, the agent constructs an ordered list of targets and delegates them accordingly.

\paragraph{\textbf{Search Agent.}} 
The \textit{Search Agent} is responsible for retrieving the goal that best matches the semantic intent extracted by the \textit{Triage Agent}, using a two-stage process that combines vector similarity search with LLM-based candidate selection. Given a target $\tau = \{\texttt{category},\; \texttt{semantics},\; \texttt{description}\}$, the agent encodes the \texttt{semantics} field using the same sentence-transformer model employed during BIM preprocessing, as described in \cref{subsec:bim-ifc}. This yields a query embedding $\mathbf{q} \in \mathbb{R}^{d}$. A cosine similarity search is performed between the query embedding $\mathbf{q}$ and all stored BIM embeddings $\mathbf{v}_i$, retrieving the top-$N$ most semantically similar candidates. For each candidate, the system computes the distance $d_i$ between the user's current position, which is estimated via the VPS, and the BIM location of the retrieved entity. The resulting set of top-$N$ candidates, denoted as $C = {(\mathbf{m}_j, d_j)}_{j=1}^{N}$, where each candidate comprises the metadata and its distance to the user, is passed to the LLM for final goal selection. The LLM is then prompted with the target $\tau$, the set of candidates $C$, and the instruction to select the goal that best aligns with the user's query, as illustrated in \cref{fig:rag-sample}. This design allows the system to resolve qualifiers such as "nearest", "closest", or "farthest", by providing the distance information in the prompt.  If no suitable match is found, the \textit{Search Agent} prompts the user for additional clarification and transfers control back to the \textit{Triage Agent} for further dialogue and orchestration. 

\paragraph{\textbf{Response Agent.}}
The role of the \textit{Response Agent} is to synthesize and generate an appropriate, context-aware conversational response. Depending on the query's category, the \textit{Response Agent} provides either direct navigation instructions, detailed informational responses, or conversational acknowledgment. This agent utilizes the structured output with the retrieved location details from previous agents, to generate concise and user-friendly  instructions.

\begin{figure}[t]
    \centering
    \includegraphics[width=0.95\linewidth]{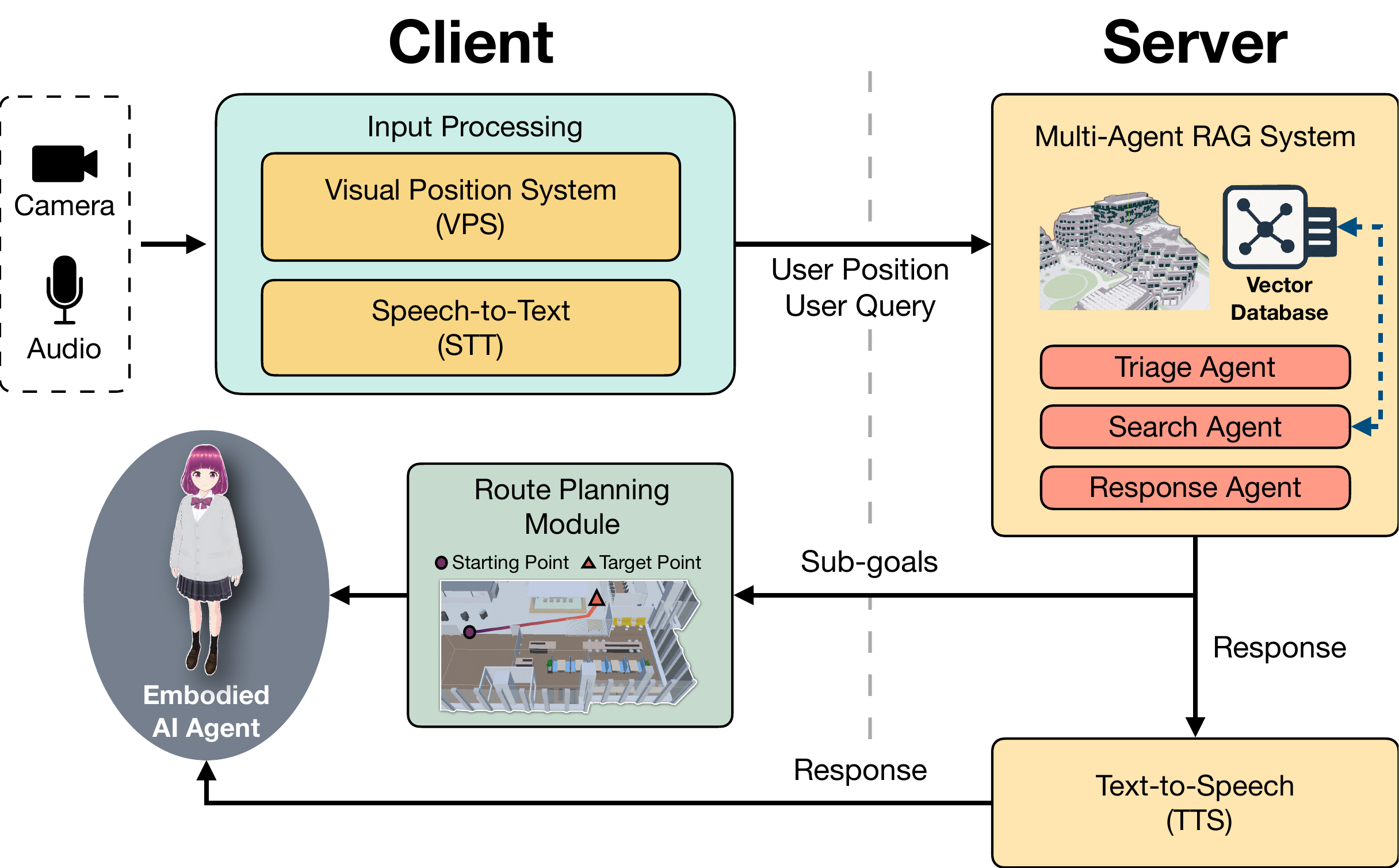}
    \caption{The overview of system design with server-client architecture. The client-side Unity application captures camera and audio input to obtain user position and query, which are sent to the server via a REST API. The server processes the query using a multi-agent RAG system powered by GPT-4o, retrieves goals from the vector database, and generates instructions and audio guidance. The selected goal is returned to the client for path planning using a pre-baked BIM-based NavMesh. An embodied AI agent then delivers AR and audio navigation guidance.}
    \label{fig:system-design}
\end{figure}

\section{System and Embodied Agent Implementation}
\label{sec:system-design}

\subsection{System Design with Server-Client Architecture}
\Cref{fig:system-design} illustrates the proposed system and the communication  between the components described in \cref{sec:multi-agent-rag}. The system is implemented through two parts: (a) a client-side Unity~\cite{unity-eng} application deployed on an iOS device to render the AR agent, perform route planning, and facilitate speech interaction, and (b) a backend server with RESTful APIs to operate the multi-agent RAG system, access to the vector database, and perform GPU-intensive operations such as inference with a high-quality text-to-speech (TTS) model.

The client-side application executes on an iPhone 15 Pro and serves as the primary interface for user interaction. It streams RGB images and inertial measurement unit (IMU) data as input. During the initialization stage, the phone's orientation and location are estimated using Niantic Lightship VPS~\cite{niantic_ardk}. The obtained orientation and location are then transformed via the pre-computed rigid transformation to align with the BIM coordinate space, as described in \cref{subsec:navigation-route-planning}. The converted pose is used to render AR content, including the virtual agent’s locomotion and navigational guidance locally in Unity application. When the user issues a voice query, speech-to-text (STT) transcription is performed using the native iOS API to convert audio into text. The transcribed query, along with the user's current location, are sent to the server to initiate the multi-agent orchestration for further processing.

On the server side, we preprocess the BIM metadata using the sentence-transformer model \texttt{"all-mpnet-base-v2"}~\cite{reimers-2019-sentence-bert}, and store the resulting embeddings with associated metadata in a Milvus vector database~\cite{2021milvus}, which supports efficient vector similarity search. At runtime, the received user query and location are forward to the proposed multi-agent RAG pipeline. We employ GPT-4o~\cite{hurst2024gpt} as the reasoning model for the \textit{Triage Agent}, \textit{Search Agent}, and \textit{Response Agent} to support natural language understanding and response generation. The same sentence encoder is used to encode the semantic keywords extracted by the \textit{Triage Agent} for semantic searches. To generate spoken guidance, we employ the MeloTTS model~\cite{zhao2024melo} for TTS synthesis, and transmit the resulting audio output back to the client in real time.

\subsection{Route Planning and Coordinate Alignment}
\label{subsec:navigation-route-planning}
Given the goal retrieved from the multi-agent RAG system and the user's current location obtained via the VPS, our planner module computes an optimal navigation route based on the BIM data. A key consideration in this process is that VPS-derived coordinates are typically defined relative to the origin of the AR system, and thus are not inherently aligned with the coordinate system of the BIM model. To enable spatial localization and route planning within the BIM environment, an additional alignment step is required to convert between the two coordinate systems. In order to compute the necessary rigid transformation, we manually align the AR anchors to the corresponding location in the BIM model. This process produces a transformation matrix $\mathcal{T}_{\text{VPS} \rightarrow \text{BIM}}$, which maps from the VPS coordinate to the BIM coordinate. During inference time, we apply the obtained transformation $\mathcal{T}_{\text{VPS} \rightarrow \text{BIM}}$ to the estimated camera pose from VPS, once it reports reliable tracking. The process is formulated as follows,
\begin{equation}
\mathcal{E}_{BIM} = \mathcal{T}_{\text{VPS} \rightarrow \text{BIM}} \cdot \mathcal{E}_{VPS},
\end{equation}
where $\mathcal{E}_{VPS}$ and $\mathcal{E}_{BIM}$ denote the camera pose in the VPS and BIM coordinate systems, respectively. The converted camera pose $\mathcal{E}_{BIM}$ enables spatial localization for AR content rendering and subsequent navigation route planning within the BIM environment. For route planning, we utilize Unity's NavMesh Agent system to construct navigation meshes from the BIM geometry and compute the shortest traversable route to the destination.

\subsection{Embodied Agent Interface for Navigation}
\label{subsec:embodied-agent}
To facilitate intuitive guidance, our proposed system employs an embodied virtual agent as a navigation interface. This humanoid agent combines voice interaction, lifelike locomotion, context-aware gestures, and adaptive user synchronization to provide guidance in a natural manner. The following paragraphs describe the details of each component.

\paragraph{\textbf{Text-to-Speech and Lip Synchronization.}}  
The agent provides verbal navigation cues through an integrated TTS model, enabling it to speak generated instructions in real time. A lip-sync mechanism is used in tandem with TTS so that the agent’s mouth movements and facial expressions align with the spoken words.

\begin{figure}[t]
    \centering
    \includegraphics[width=0.95\linewidth]{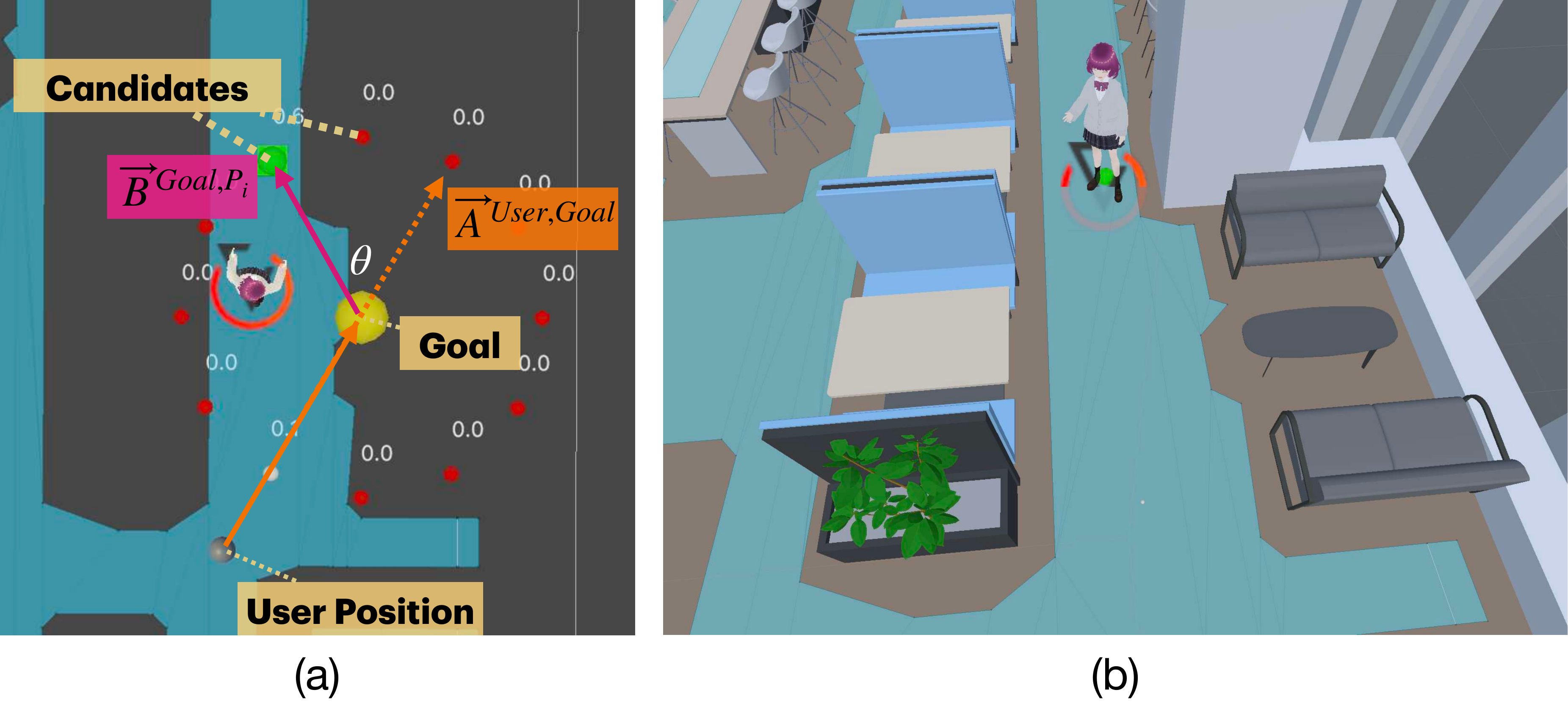}
    \caption{The illustration of the side-stepping mechanism: (a) Ten candidate positions $P_i \in \mathcal{P}$ are sampled around the destination. Each candidate is scored based on the angle $\theta$, computed between vectors $\vec{A}^{\text{User}, \text{Goal}}$ and $\vec{B}^{\text{Goal}, P_i}$. Smaller angles are favored, as they align the agent to face both the user and the goal. (b) Example of a selected final position.}
    \label{fig:side-stepping}
\end{figure}
\paragraph{\textbf{Locomotion with Blend Trees and Animation Layering.}} 
To achieve smooth and responsive movement, the agent’s animation is implemented using Unity’s Blend Tree system, which interpolates between locomotion states  based on the character’s speed and orientation. Upper-body gestures are managed through animation layering, allowing the agent to perform communicative actions, such as pointing or speaking, while walking. This layered structure enhances expressiveness without disrupting locomotion.

\paragraph{\textbf{Side-Stepping for Presentation Pose Selection.}}
To improve the believability of how the character concludes its navigation at a destination, we introduce a technique termed side-stepping. Instead of stopping exactly at the target point, the embodied agent takes a small step to the side, enabling it to turn and face the user while gesturing toward the destination, creating a more natural and communicative stance. This behavior is implemented by sampling ten candidate positions arranged in a circular pattern around the destination. Each position is scored based on two criteria: (1) angular alignment relative to the user’s position, favoring locations that orient the character toward both the user and the destination, and (2) walkability, determined by checking NavMesh availability at each point. The character transitions to the highest-scoring position shortly before reaching the original destination, which allows for a clean and intentional presentation gesture. An example of the sampling and selection process is illustrated in \cref{fig:side-stepping}. 

\paragraph{\textbf{Adaptive Wait-Resume Mechanism.}} 
To remain in sync with the user’s walking pace, the agent incorporates an adaptive waiting mechanism based on two distance thresholds: (a) the wait threshold $\tau_{\text{wait}}$ and (b) the resume threshold $\tau_{\text{resume}}$. If the user falls behind and the distance exceeds $\tau_{\text{wait}}$, the agent gradually slows down and transitions into an idle or waiting state. In this state, the agent turns partially toward the user and adopts a relaxed stance, signaling that it is pausing. The agent does not resume walking until the distance between the user and the agent falls within $\tau_{\text{resume}}$. By using separate wait and resume triggers, the system avoids frequent toggling of the walk animation in response to minor distance fluctuations, thereby ensuring stable synchronization.

\section{Experimental Results}
\subsection{Research Objectives of User Study}
The primary objective of this user study is to understand what kind of interface best complements the intelligence and flexibility afforded by the proposed multi-agent and RAG-based system for AR navigation. While the proposed RAG system enables dynamic and semantically rich query handling, the user experience ultimately hinges on how this intelligence is presented, through a human-like virtual agent or minimalist cues. To guide this investigation, we pose two research questions (RQs) as follow:
\begin{itemize}[noitemsep]
\item \textbf{RQ1:} Which interface modality, either an AR embodied agent or AR arrows, best complements the intelligence and flexibility of the proposed navigation system as evidenced by user perceptions of system intelligence, coherence, and trust?
\item \textbf{RQ2:} How usable is the proposed multi-agent and RAG-based system for goal retrieval when experienced through the AR embodied agent interface in the navigation task?
\end{itemize}

To address these questions, we first designed a comparative user study to evaluate two AR guidance modalities: (a) AR directional arrows, and (b) an embodied virtual agent overlaying the same AR arrows within a fully integrated navigation system. This system combines BIM data for spatial awareness, flexible natural language queries for goal specification, and a multi-agent RAG system to interpret user queries and reason over spatial-semantic information extracted from the environment. While both guidance approaches have been widely used in AR navigation, prior research has identified trade-offs between them. AR arrows offer a minimalist and familiar visual cue but may struggle to convey complex spatial transitions or semantic intent. In contrast, agent-based guidance can improve spatial legibility and perceived presence, fostering a more intuitive and engaging experience, yet may increase cognitive load or visual clutter in constrained environments. Mixed reality studies have shown that both modalities are consistently ranked among the most preferred interfaces due to their simplicity and clarity, indicating that a direct, controlled comparison remains valuable. To this end, our study pursues two complementary aims: (1) to determine which interface modality best complements the intelligence and flexibility afforded by the proposed system, and (2) to assess the overall usability of the system when experienced through the embodied agent interface. We employ the System Usability Scale (SUS) to assess the overall usability of the proposed system, as perceived by users when interacting with the embodied agent interface.

\subsection{User Study Design}
We employed a within-subjects study design, with the guidance modality as the independent variable. Each participant experienced both scenarios of the AR navigation system: one utilizing directional AR arrows for guidance, and the other employing an embodied agent complemented by the same arrow cues. A within-subject design was chosen to control for individual differences in navigation ability and technology experience, and to allow participants to directly compare the two modalities. Each participant will perform two navigation tasks in each condition, as detailed in the following paragraphs. Throughout the study, the dependent variables collected will include both objective performance metrics (e.g., success rate and errors) and subjective evaluation metrics (e.g., perceived trust, engagement, etc.), described in the \cref{subsec:evaluation-metrics} section.

\subsubsection{Participants}
We invited $N = 20$ participants to take part in the study, representing a mix of genders and a diverse age range (approximately 20–50 years old). No specialized knowledge of augmented reality (AR) or technical background was required; instead, we sought individuals with typical familiarity using smartphones or navigation applications. All participants were screened for normal or corrected-to-normal vision and functional proficiency in English, as both the AR interface and associated questionnaires were presented in English. Prior to participation, individuals provided informed consent and completed a brief pre-study questionnaire to collect demographic information and assess any prior experience with AR technologies. During the study, participants were instructed to interact with the AR system as they would with a conventional navigation aid, performing the tasks naturally and without overthinking the interface.

\subsubsection{Evaluation Metrics}
\label{subsec:evaluation-metrics}
To comprehensively evaluate the AR navigation system, we collected both objective and subjective metrics. In particular, we adopt a within-subjects comparative evaluation framework in which participants experience and directly compare two guidance scenarios: (1) AR arrows alone, and (2) an embodied agent guide augmented with the same arrow cues. After completing both tasks, participants responded to a series of post-task questionnaires using 5-point Likert scale items, allowing side-by-side comparison across modalities. In addition, the System Usability Scale (SUS) score is applied to evaluate the overall usability of the system in the embodied agent condition. The details of the applied measures are described below,

\metricitem{Task Success Rate.} 
A task is considered successful if the participant reaches the destination independently without requiring assistance. Failures are recorded when the participant is unable to reach the destination or must retry due to an unsuccessful retrieval. These failures are categorized based on their underlying causes, which include (1) STT transcription errors, (2) retrieval or reasoning failures from the RAG system, and (3) invalid operations, such as accidental button presses or mis-interactions with the system interface.

\metricitem{Goal Retrieval Rate.}  It is defined successful if the participant is able to obtain the correct navigation goal from a language query.

\metricitem{Perceived Workload.}  To assess cognitive demand, participants are asked to rate their perceived workload. This measures whether the navigation scheme imposes greater cognitive burden than the other.

\metricitem{System Usability Scale (SUS).}  The SUS~\cite{bangor2009determining, lewis2018item} is used to evaluate the overall usability of the agent interface. It produces a score ranging from 0 to 100, with higher scores indicating better usability.

\metricitem{Perceived Trust.} Participants assess the level of trust they place in each guidance modality with a comparative question: “Which guidance felt more trustworthy or confident?” This evaluates whether the embodied agent enhances users’ navigational confidence.

\metricitem{Engagement and Enjoyment.} Participants respond to a question which compares two modalities in terms of engagement: “Which method was more engaging or enjoyable?” This metric captures the motivational dimensions of the user experience.

\metricitem{Perceived Intelligence.} This measure reflects how intelligent and responsive the proposed system appears in each scenarios. It is particularly relevant in the agent scenario, where the embodied agent conveys intelligence through both natural language interaction and locomotion; while in the arrow-only scenario, participants may perceive the system's intelligence as stemming primarily from the  RAG functionality, which serves as a baseline for comparison.

\metricitem{Perceived Clarity and Comfort.} This metric helps interpret usability differences: while arrows may be clean but ambiguous, agents may enhance clarity or introduce new forms of discomfort.

\subsubsection{Experiment Procedure and Tasks}
In the study, participants are asked to interact with the system via natural language voice commands to specify their destination (goal) \textbf{\textit{"in their own words"}} (e.g., “Take me to the cafeteria”). The proposed multi-agent RAG strategy, as described in \cref{subsec:multi-agent-rag}, is employed to answer contextual queries by retrieving relevant information from the vector database and generating user-friendly responses. Each participant is asked to complete two navigation tasks under each guidance scheme, resulting in four tasks in total. For each task, the user starts from a common location and follows the AR navigation guidance to a designated target within the building. Target destinations vary across tasks and include rooms and facilities such as the “Reception” or “Largest Meeting Room.” To prevent participants from memorizing routes or retracing prior paths, distinct destinations are used in each trial, ensuring that every task remains novel and representative of real-world navigation challenges. Each study session takes approximately 10–12 minutes, including a brief tutorial at the beginning (1–2 minutes) and four navigation tasks (8–10 minutes). The questionnaire is answered once at the end of the session after completing all tasks under both conditions.

\begin{figure}[t]
    \centering
    \includegraphics[width=\linewidth]{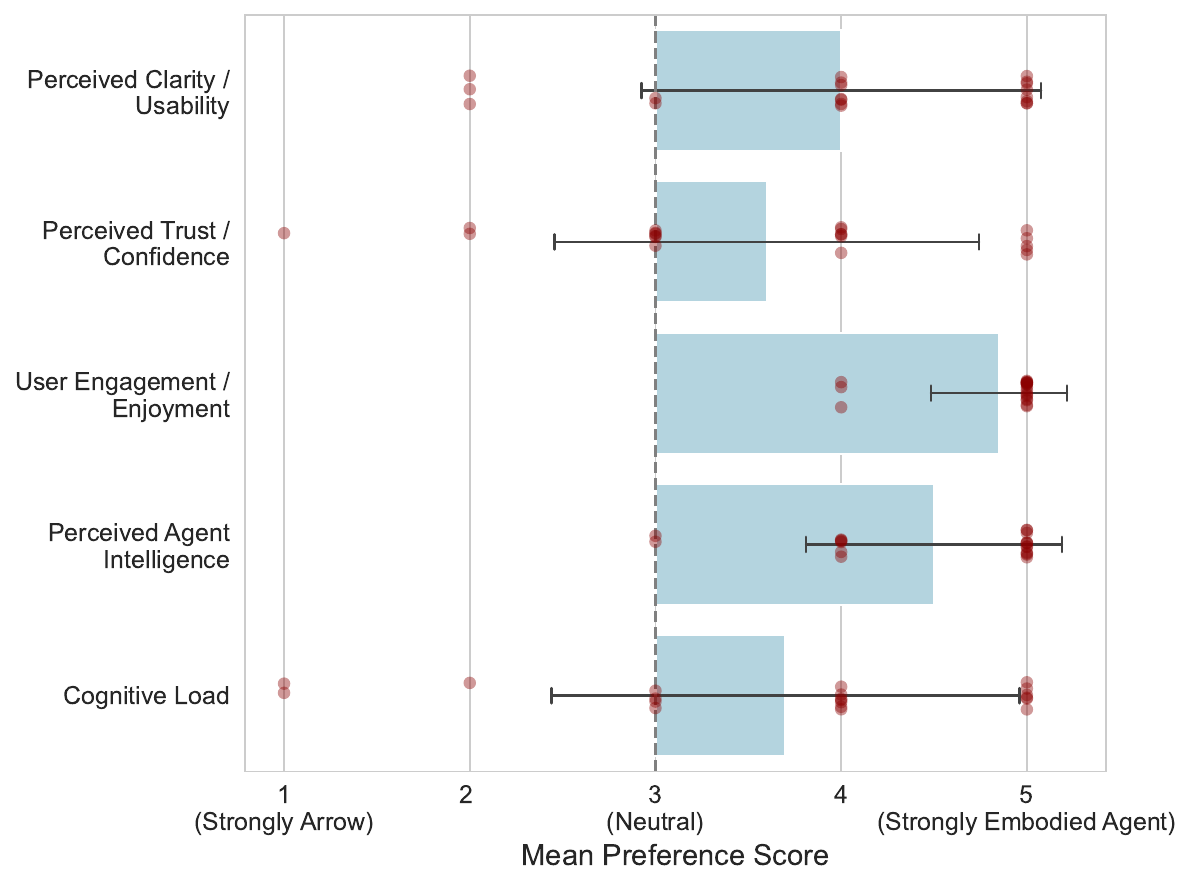}
    \caption{Comparative evaluation of user preferences between arrow-only and embodied agent interfaces. Positive values indicate preference for the embodied agent, while negative values reflect preference for the arrow-only scheme. Error bars represent standard deviation, and individual sample points are shown as scatter dots.}
    \label{fig:comparative-figure}
\end{figure}

\subsection{Comparative Evaluation on Navigation Interfaces}
\label{subsec:user-exp}
The results of the user experience is presented in \cref{fig:comparative-figure}. The analysis of participant responses ($N = 20$) reveals clear preferences across several measures. A Wilcoxon signed-rank test is performed to compare participant ratings against ratings against the neutral midpoint value of $3$ on a $1$–$5$ Likert scale. Effect sizes are reported using the rank-biserial correlation $r$, which quantifies the consistency and magnitude of directional preference, where higher absolute values indicate stronger effects~\cite{Kerby2014TheSD}.

In this study, participants consistently rate the embodied agent interface higher than the arrow-only condition in terms of perceived clarity and usability ($M = 4.0$, $r = 0.81$, $p = .002$), engagement and enjoyment ($M = 5.0$, $r = 1.00$, $p < .001$), and perceived intelligence ($M = 5.0$, $r = 1.00$, $p < .001$). All three effects are statistically significant and reflect large effect sizes. Engagement receives the highest ratings with all participants rating it above the neutral point, which reflects complete agreement in the direction of preference. This suggests that participants find the agent guidance significantly more enjoyable and interactive than the arrow baseline.  Perceived intelligence also shows a strong effect and preference toward embodied agent interface, suggesting that users view the agent-equipped RAG system as more intelligent and context-aware. It is worth noting that, both guidance modalities employ the same underlying RAG system; however, the agent interface is rated significantly higher. This result highlights a key insight: \textit{\textbf{the presence of an embodied agent, equipped with voice interaction and locomotion, can meaningfully amplify users’ perception of system intelligence, even when the reasoning backend remains identical.}} This suggests that embodiment plays a critical role in shaping user impressions of intelligence and responsiveness in AR systems.

While the most prominent differences are observed in clarity, engagement, and perceived intelligence, participant responses also reveal statistically significant trends in trustworthiness and cognitive load. The difference in trustworthiness ratings relative to the neutral point is moderate ($M = 4.0$, $r = 0.61$, $p = .038$), reflecting a statistically significant but smaller effect. Cognitive load is rated at the neutral midpoint with a modest positive shift in favor of the agent interface ($M = 4.0$, $r = 0.57$, $p = .040$). While the effect is statistically significant, the median score suggests that the agent interface does not substantially alter users' mental effort compared to the baseline approach, and participants tend to find it marginally easier to interpret and follow.

\begin{figure}[t]
    \centering
    \includegraphics[width=\linewidth]{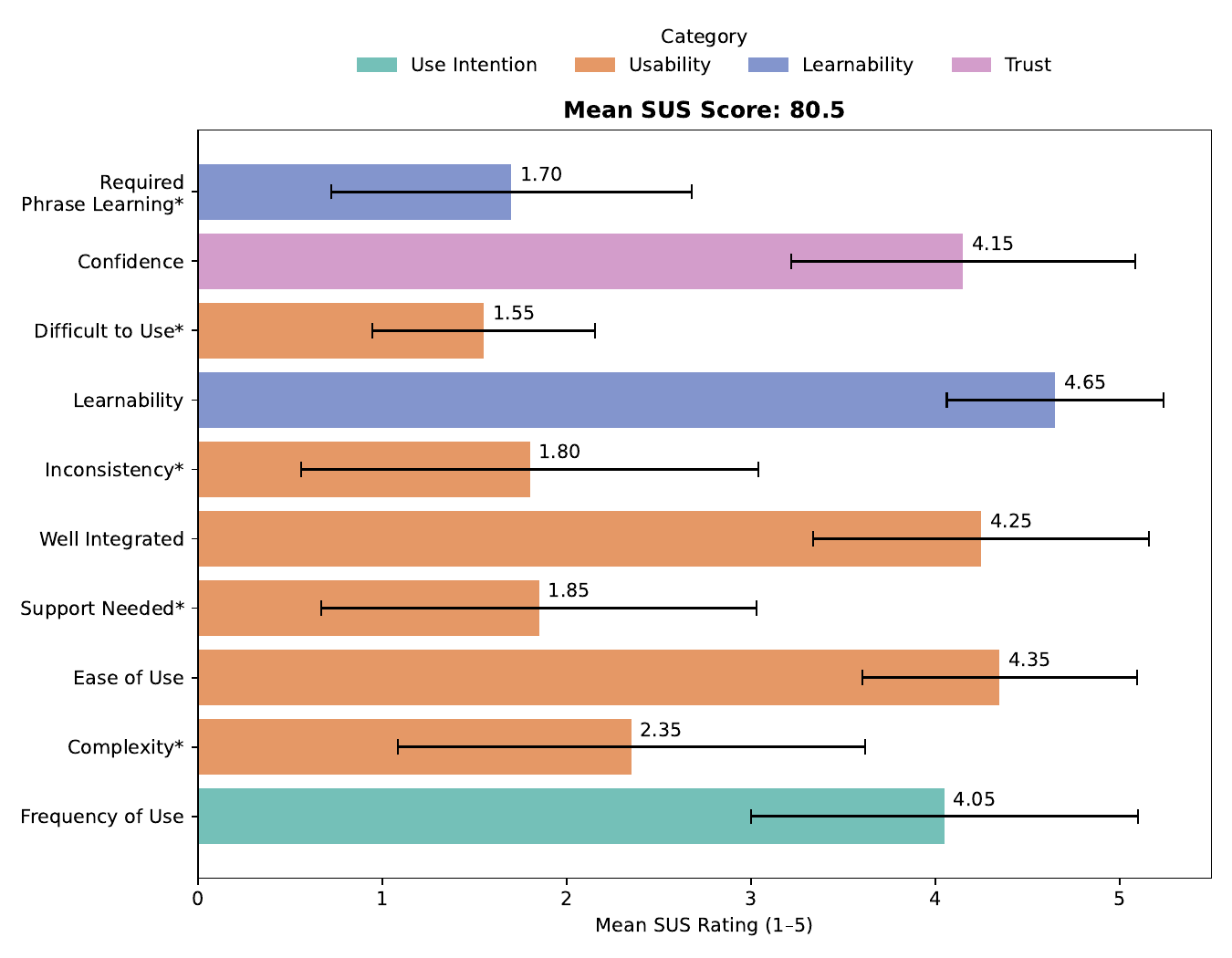}
    \caption{The system usability scale (SUS) results for the proposed multi-agent RAG system with an AR embodied agent. Labels marked with an asterisk (*) indicate reverse-scored items, where lower raw scores reflect more favorable responses. The SUS score is calculated using standard scoring procedures, which results in a composite usability rating out of 100. Error bars represent standard deviation.}
    \label{fig:sus-figure}
\end{figure}

\subsection{System Usability Evaluation}
\Cref{fig:sus-figure} presents the results of the System Usability Scale (SUS) score assessment. The proposed system yields a mean score of 80.5 (SD = 11.5) out of 100, which indicates a high level of usability. This outcome is consistent with established SUS benchmarks, where scores above 80 are interpreted as “excellent” and comparable to highly rated interactive systems~\cite{bangor2009determining, lewis2018item}. The analysis of individual SUS items reveals that participants rate the system favorably in terms of ease of use (M = 4.35, SD = 0.75), learnability (M = 4.65, SD = 0.59), and the integration of system features (M = 4.25, SD = 0.91). Participants also express confidence in using the system (M = 4.15, SD = 0.93) and indicate a clear intention to use it frequently (M = 4.05, SD = 1.05). Perceived complexity receives a slightly lower rating (M = 2.35, SD = 1.27), suggesting that users perceive the interaction process to be moderately cumbersome or less streamlined. Participants attribute these issues primarily to the requirement of pressing and holding a button during voice input. Overall, the SUS results demonstrates that the embodied agent interface exhibits a high level of usability and supports a positive user experience. When considered alongside the comparative findings described in~\cref{subsec:user-exp}, the results further support the conclusion that the overall system design contributes meaningfully to user engagement, trust, and perceived system intelligence.

Following the SUS evaluation, we analyze system-level performance to understand the operational reliability of the system. The overall task success rate is reported as 89.8\%, which indicates that the vast majority of participants are able to reach their intended destination without external assistance. In addition, the goal retrieval success rate reaches 97.7\%, suggesting that the system is highly effective at interpreting natural language queries and identifying the correct target destination that matches user's intention. We further analyze the root causes of failure cases to gain insight into specific system limitations and user interaction issues. The remaining 10.2\% of system-level failures can be attributed to two primary sources, which include STT transcription errors (3.4\%) and invalid operations or input failures (4.5\%). The former includes cases where the STT module fails to recognize the user’s query or mis-transcribes it, which leads to failed goal retrieval attempts because the system interprets the query as referring to non-existent or invalid destinations. The latter includes instances where users unintentionally mis-pressed the voice interaction button or issued fragmented inputs, thereby hindering successful execution. The remaining errors consist of goal retrieval failures (2.3\%) and other errors (1.1\%) caused by gender-specific queries. We further provide case studies of both successful and failed scenarios in~\cref{subsec:case-study} to analyze the system’s behavior and limitations in depth.

\begin{figure}[t]
    \centering
    \includegraphics[width=\linewidth]{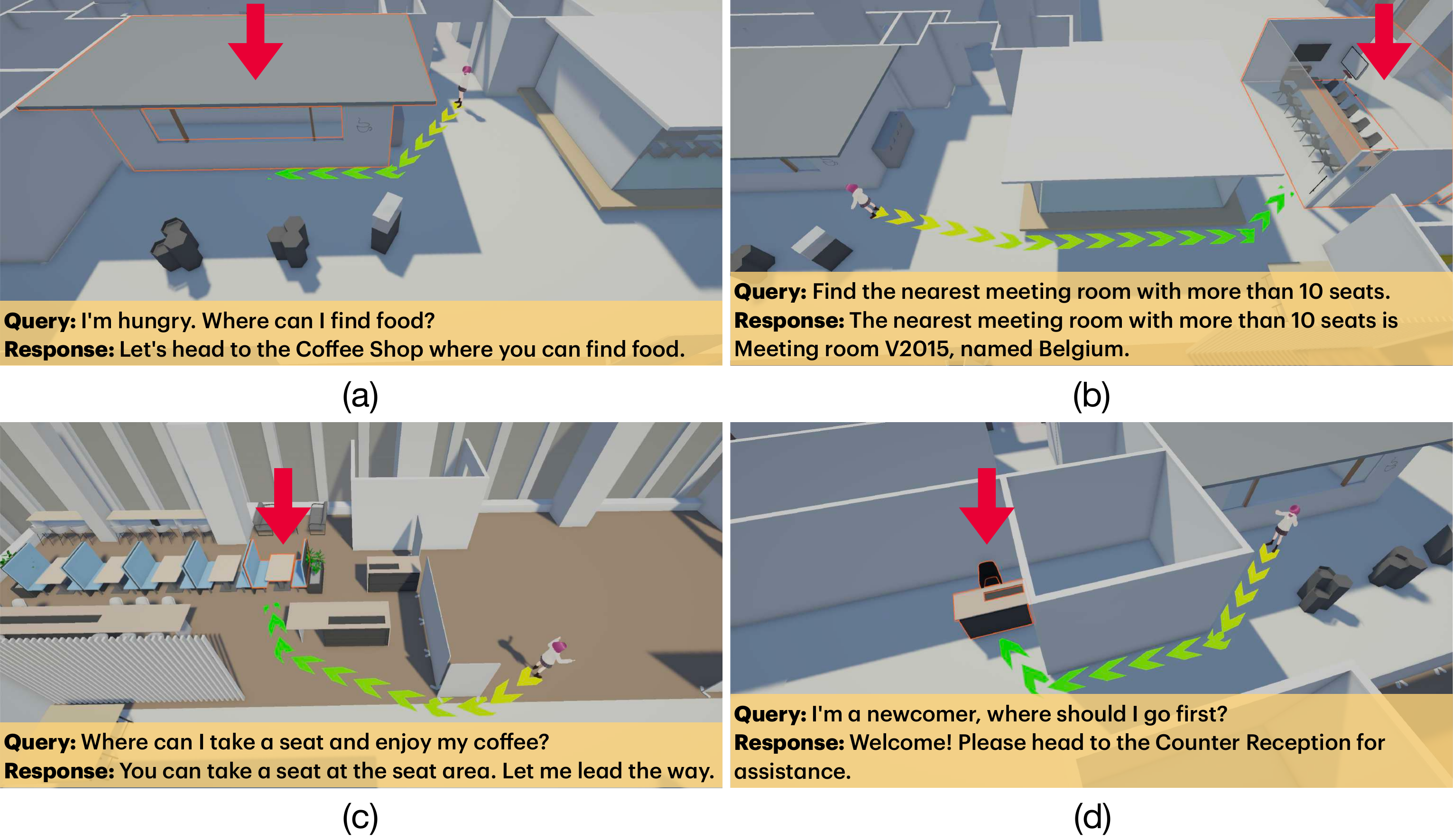}
    \caption{Examples of successful cases of goal retrieval and navigation route generation with the proposed multi-agent RAG system, which demonstrate its ability to interpret language queries in context.}
    \label{fig:query-sample}
\end{figure}

\subsection{Case Study of Goal Retrieval with Multi-Agent RAG}
\label{subsec:case-study}
To further evaluate the proposed system's behavior, we present a set of case analyses on user queries and the corresponding responses. These examples highlight both successful and failure cases, offering insights into the strengths and limitations of the RAG-based goal retrieval. The observations are detailed in the following paragraphs, and \cref{fig:query-sample} illustrates the successful cases of goal retrieval.

\metricitem{Gender-Specific Destination Inference.}
In some cases, the system fails to infer user-specific needs when not explicitly mentioned. For example, when a female user asked, \textit{"Hello, can you take me to the toilet"}, the agent responded, \textit{"Sure, I'll guide you to the Men's Toilet"}, demonstrating a lack of gender-awareness unless explicitly stated. However, when the user added gender-specific information, such as \textit{"Hey can you take me to the women's toilet"} or \textit{"Where is the restroom. I'm a girl."}, the agent successfully guided the user to the correct facility. This case reveals the importance of explicit user context in language queries, which could be mitigated by incorporating user profile information.

\begin{figure*}[t]
    \centering
    \includegraphics[width=\linewidth]{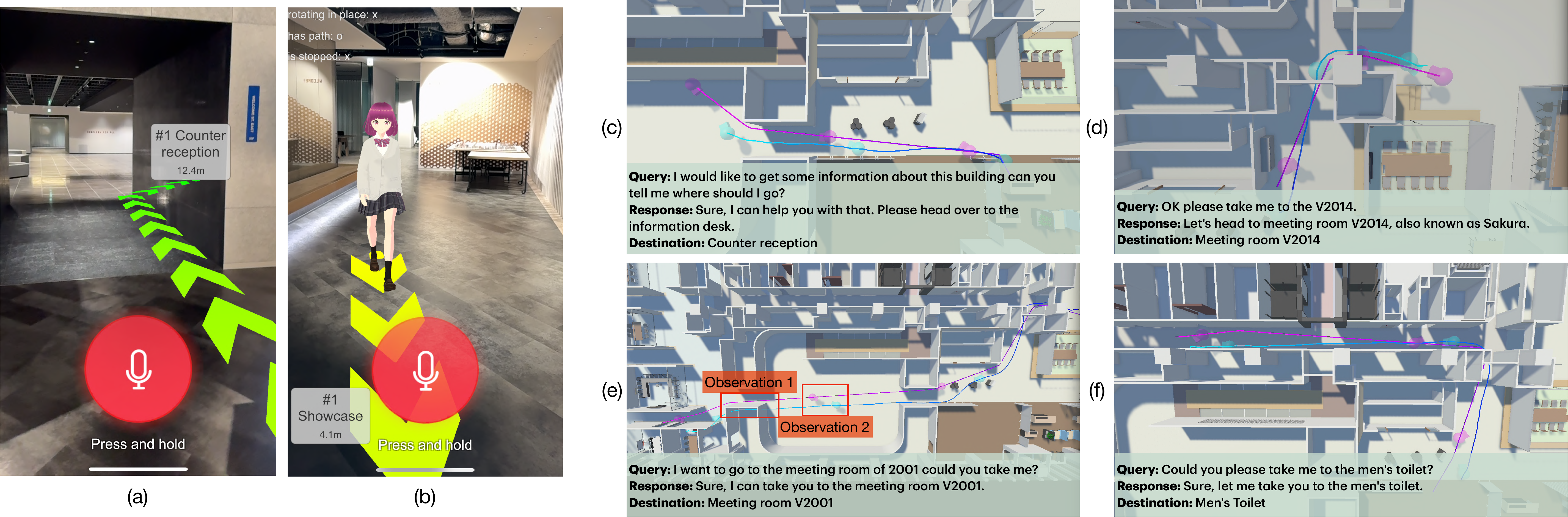}
    \caption{\textbf{Left:} Illustrations of the AR navigation schemes, including (a) the arrow-only scheme, and (b) the proposed embodied agent scheme. \textbf{Right:} Visualizations of user trajectories and the navigation paths traversed by the embodied agent. The purple lines represent the planned navigation routes followed by the agent, while the aqua blue lines indicate the actual walking trajectories of users.}
    \label{fig:qualitative-results}
\end{figure*}

\metricitem{Semantic Confusion Between Nearby Locations.}
The system occasionally misidentifies the intended destination when semantically similar or spatially adjacent options exist. In one instance, a user requested, \textit{"Please take me to Café"}, but the system navigated the user to the coffee shop seating area instead of the actual coffee shop. This behavior suggests that the underlying sentence transformer-based encoder~\cite{reimers-2019-sentence-bert,reimers-2020-multilingual-sentence-bert} may prioritize surface-level similarity over a deeper semantic understanding of the intended location. As a result, the system selects a nearby but unintended place. Incorporating more advanced models with stronger semantic representation or reasoning capabilities could mitigate these types of errors.

\metricitem{Spatial Reasoning with IFC-Derived Metadata.}
The proposed multi-agent RAG system demonstrates strong capability in interpreting queries with spatial information and context. When asked, \textit{"Can you take me to the largest meeting room?"}, the agent was able to successfully identify the largest room based on IFC data, and responded with, \textit{"Sure, let me guide you to the largest meeting room V2003. It is 40 square meters and has a seating capacity of 12."} This showcases the potential of combining semantic language understanding with BIM data.

\metricitem{Context-Aware Query and Response.}
As shown in the example in~\cref{fig:query-sample}~(a), the proposed system is capable of processing indirect and contextual queries to retrieve relevant goals. For instance, given the query "I'm hungry, where can I find food?", the agent correctly interprets the user’s intent and retrieves the goal "coffee shop." Beyond navigation, the system is also able to respond to non-navigational questions. For example, the query \textit{"When does the coffee shop open?"} received the response, \textit{"The coffee shop is open from 11:00 AM to 6:00 PM."} These examples highlight the advantages of multi-agent orchestration and demonstrate the system’s utility as an informational agent capable of delivering context-aware responses.

\metricitem{Speech Recognition Errors and Ambiguity.}
Some failure cases are observed due to incorrect STT transcription, particularly among non-native English speakers. For instance, when a user intends to ask for the "men's toilet," the utterance is transcribed as \textit{"Where is the main toilet"}, resulting in a query the agent cannot resolve. Similarly, \textit{"Can you get me to the 201 for meeting room"} is misinterpreted, which leads to failure in retrieving the correct destination (e.g., meeting room 2014). These cases highlight the need for more robust STT handling and fallback mechanisms when processing voice-based queries with uncertain or ambiguous input.


\subsection{Qualitative Results}
\label{subsec:qaulitative-results}
In this section, we provide additional qualitative results, including (a) the screenshots of the interfaces for both the arrow-only and agent-guided scenarios, and (b) visualizations of users' walking trajectories and the planned navigation routes of the embodied agent, as shown in \cref{fig:qualitative-results}. To record user positions during navigation, we capture poses using a combination of the VPS and IMU data, and the results are depicted in \cref{fig:qualitative-results}~(c)-(f). Across all samples, users are able to successfully reach their destinations under the guidance of the embodied agent. It can be observed that the spatial distance between the user and the agent remains relatively close throughout the navigation process, as maintained by the adaptive waiting mechanism described in \cref{subsec:embodied-agent}. In \cref{fig:qualitative-results}~(c), we observe that the proposed system is capable of retrieving the correct navigation goal even when the user’s natural language query is indirect (e.g., referring to the target contextually rather than by name). In \cref{fig:qualitative-results}~(e), we highlight two notable observations. First, in \textbf{\textit{Observation~1}}, part of the user's trajectory (shown in aqua blue) appears to intersect with walls. This is attributed to drift in the position estimation, where accumulated IMU error after the last VPS correction leads to the inaccurate positioning. This issue could potentially be mitigated by integrating additional AR anchors to stabilize pose estimation. Second, in \textbf{\textit{Observation~2}}, the agent is observed to stop and orient itself toward the user while waiting for the user to catch up, which demonstrates the effectiveness of the proposed adaptive waiting mechanism described in \cref{subsec:embodied-agent}.


\section{Discussion}
\subsection{Reflections on Research Questions}
Our findings provide key insights into RQs and contribute to the design of embodied AR navigation systems integrated with RAG-based reasoning. The results show that the virtual agent significantly enhances perceived system intelligence and coherence, even though both modalities rely on the same multi-agent RAG framework. These outcomes suggest that embodiment serves as a key mediator, helping users interpret system behavior as more intelligent and context-aware. The presence of a human-like agent appears to manifest the system’s reasoning to the user, making its actions feel more intelligent and responsive. In contrast, measures of user trust and cognitive load show no substantial difference between the two modalities. Based on user comments and case study observations, we observe that while the virtual agent contributes positively to perceived intelligence, its behavioral realism remains limited. The current implementation employs a fixed set of animations that do not fully adapt to context or user interaction. Moreover, the moderate quality  and fidelity of AR rendering may also affect the virtual agent’s overall believability, thereby constraining its potential to enhance user trust. On the other hand, the requirement to hold the phone during the navigation process may introduce physical effort, particularly for users who are less familiar with handheld smartphone interactions, which in turn could increase cognitive load.  Future iterations of the system may benefit from more realistic animation, gestural responses, and hands-free delivery using AR glasses. Such improvements could enhance both trust and ease of use, while further amplifying perceived intelligence and user experience.

\subsection{Limitation and Future Direction}
The proposed system demonstrates the potential and promising utility of integrating multi-agent RAG system with spatial metadata derived from BIM and IFC. However, several aspects of the current implementation suggest directions for future research. First, the system currently assumes the presence of BIM and IFC data, which serves as the foundational source for spatial and semantic data. While BIM enables precise path planning and contextual goal retrieval, this requirement may limit deployment in environments where such models are unavailable or incomplete. Additionally, BIM data may not always reflect the most recent physical state of the environment, e.g., furniture may have been moved or spaces reconfigured, potentially leading to mild spatial misalignment in rendered guidance. Another consideration is the reliance on visual positioning for aligning AR content with the physical space. As noted in our  observations in \cref{subsec:qaulitative-results}, location drift or misalignment can occur due to limitations in camera-based localization or calibration discrepancies between the BIM and AR anchor systems. 

To broaden the system’s applicability in environments lacking BIM or IFC data, a promising future direction involves incorporating scene reconstruction techniques alongside semantic understanding. Methods such as TSDF fusion~\cite{newcombe2011kinectfusion, zeng20163dmatch} with LiDAR, or learning-based approaches~\cite{sayed2022simplerecon, sayed2024doubletake}, could enable the generation of spatial geometry. To enrich this geometry with semantic metadata, recent advances in 3D semantic instance segmentation~\cite{takmaz2023openmask3d, boudjoghra2025openyolo} can be leveraged to detect and classify common indoor objects. While attaching detailed descriptions to the detected objects remains an open challenge, such methods provide a strong foundation for enabling richer scene understanding. For practical deployment, one viable pathway is to develop an interactive tool that allows users to customize and annotate reconstructed scenes with object labels or descriptions through 3D bounding boxes. These annotations can be serialized into IFC structures, providing a lightweight means to update or construct usable semantic maps for navigation tasks. In parallel, improving the spatial alignment between the BIM model and the AR environment remains an open challenge. Current implementations rely on manual calibration between BIM and VPS coordinate systems, which can be time-consuming and error-prone. This process may be accelerated through automated registration methods that estimate correspondences between BIM and AR anchors.

\section{Conclusion}
We presented an AR navigation system that integrates a multi-agent RAG framework with BIM data to support flexible and language-driven goal retrieval. The system allows users to specify navigation targets through natural language, which are then interpreted by a multi-agent orchestration system and visualized through an AR embodied agent interface. A user study conducted in a real-world indoor environment demonstrates the system’s effectiveness, with a SUS score of 80.5, which is indicative of excellent system usability. In addition, comparative evaluations reveal that the embodied agent, equipped with voice interaction and locomotion, significantly enhances users’ perception of system intelligence, even when the underlying reasoning remains unchanged. These findings underscore the importance of embodiment in designing intelligent and user-centered AR navigation systems.

\bibliographystyle{abbrv-doi-narrow}

\bibliography{template}
\end{document}